\documentclass[global,twocolumn,showpacs]{svjour}
\usepackage{graphicx}
\usepackage{amssymb}
\journalname{Appl. Phys. A}

\begin{document}

\title{Graphene based superconducting quantum point contacts}
\author{Ali G. Moghaddam and Malek Zareyan}

\institute{Institute for Advanced Studies in Basic Sciences
(IASBS), P. O. Box 45195-1159, Zanjan 45195, Iran}
\mail{zareyan@iasbs.ac.ir}
\date{Received: date / Revised version: date}
\maketitle
\begin{abstract}
We investigate the Josephson effect in the graphene nanoribbons of
length $L$ smaller than the superconducting coherence length and
an arbitrary width $W$. We find that in contrast to an ordinary
superconducting quantum point contact (SQPC) the critical
supercurrent $I_c$ is not quantized for the nanoribbons with
\textit{smooth} and \textit{armchair} edges. For a low
concentration of the carriers $I_c$ decreases monotonically with
lowering $W/L$ and tends to a constant minimum for a narrow
nanoribbon with $W\lesssim L$. The minimum $I_c$ is zero for the
smooth edges but $e\Delta_{0}/\hbar$ for the armchair edges. At
higher concentrations of the carriers this monotonic variation
acquires a series of peaks. Further analysis of the current-phase
relation and the Josephson coupling strength $I_cR_N$ in terms of
$W/L$ and the concentration of carriers revels significant
differences with those of an ordinary SQPC. On the other hand for
a \textit{zigzag} nanoribbon we find that, similar to an ordinary
SQPC, $I_c$ is quantized but to the half-integer values
$(n+1/2)4e\Delta_{0}/\hbar$.
\end{abstract}

\section{Introduction}
\label{sec:1} Graphene, a single layer graphite, is the two
dimensional honeycomb lattice of carbon atoms. Recently
experimental realization  of graphene has introduced a new
material with unique properties
\cite{novoselov04,novoselov05,zhang05} which provide the
possibility for designing new carbon based nanodevices. Most of
the peculiarities come from the electronic structure of graphene
which is fundamentally different from that of a metal or a
semiconductor. Graphene has a gapless semi-metallic band structure
with a linear dispersion relation of low-lying excitations. This
makes the electrons in graphene to behave identical to two
dimensional massless Dirac fermions
\cite{wallace,slonczewski,semenoff}. The low energy physics of
electrons is then described by a two dimensional Dirac Hamiltonian
acting on a four-component spinors in two sublattice spaces of the
honeycomb lattice and two independent $K$ and $K^{\prime}$ points
in the corresponding reciprocal lattice \cite{mele,ando}. As a
condensed matter counterpart of relativistic particles
\cite{haldane}, graphene makes possible to observe early
predictions like Klein paradox and zitterbewegung effect
\cite{katsnelson}. Also it has attracted an intense theoretical
and experimental attention to study the effects of the
relativistic-like dynamics of electrons on the different quantum
transport phenomena which are already known in ordinary conducting
systems.
\par
Already several quantum transport phenomena including the integer
quantum Hall effect \cite{novoselov05,zhang05,gusynin05},
conductance quantization \cite{peres06}, quantum shot noise
\cite{beenakker06-1}, and quantum tunneling \cite{katsnelson} have
been revisited in graphene and found to have anomalous features
due to the relativistic like dynamic of electrons. Recently the
possibility of the Josephson coupling of two superconductors by a
graphene layer was considered by Titov and Beenakker
\cite{beenakker06-2}. Using Dirac-Bogoliubov-de Gennes formalism
\cite{beenakker06-3}, they found that in a ballistic graphene a
Josephson current can flow even in the limit of zero concentration
of the carriers \textit{i.e.} at the Dirac point. They found a
current-phase relation
$I(\phi)\propto\cos(\phi/2){\mathrm{arctanh}}(\sin(\phi/2))$ with
critical current depending inversely on the junction length $L$.
These results for a wide ballistic graphene junction at the Dirac
point are formally identical to those of an ordinary
\textit{disordered} normal metal \cite{beenakker91-2,kulik1}.
\par
The graphene Josephson junction has been studied experimentally
very recently by Heersche \textit{et al.} \cite{heersche}, and
also independently Shailos \textit{et al.} \cite{shailos}. They
have confirmed the proximity induced superconductivity in the
graphene samples with superconducting electrods at the top of
them. Heersche \textit{et al.} have detected a Josephson
supercurrent in graphene monolayers at Dirac point with a bipolar
property due to the electron-hole symmetry \cite{heersche}.
\par
While the experiments and theory mentioned above had been focused
on the wide graphene Josephson junctions, fabrication of graphene
nanoribbons with finite widths were reported in a very recent
experiment \cite{chen07}. Already several theoretical works have
predicted important properties for graphene nanoribbons resulting
from the electronic states formed at the edges
\cite{louie1,louie2,rycerz06}. So one would expect interesting
properties for the graphene based junctions with different edges
which introduce a new class of Josephson nanodevices.
\par
In an ordinary superconducting quantum point contact (SQPC)
shorter than superconducting coherence length it was predicted by
Beenakker and van Houten \cite{beenakker91-1} that the critical
(maximum) supercurrent shows a step-like variation with width of
the contacts of order of the Fermi wavelength of electrons.
Furusaki \textit{et al.} \cite{furusaki91} considering such a
system under more general assumptions obtained the same results.
The quantization of the supercurrent resembles the normal
conductance quantization in a ballistic quantum point contact to
$2e^2/\hbar$ times the number of the transparent quantum channels
\cite{vanwees88,wharam88}. The quantization of the supercurrent in
units of $e\Delta_{0}/\hbar$ ($\Delta_{0}$ is the superconducting
gap in the electrodes) was confirmed experimentally in
semiconducting heterostructures \cite{takayanagi95}.
\par
In this paper we study the Josephson current in a graphene based
SQPC with length $L$ smaller than the superconducting coherence
length $\xi=\hbar \mathrm{v}_{F} /\Delta_{0}$, and an arbitrary
width $W$, with smooth, armchair and zigzag edges. Within the
formalism of Ref. \cite{beenakker06-3} we find that in contrast to
an ordinary SQPC \cite{beenakker91-1,furusaki91} the critical
supercurrent $I_c$ in smooth and armchair ribbons with a low
concentration of the carriers is not quantized but rather shows a
monotonic decrease by lowering $W$. For a narrow strip $W\lesssim
L$ the supercurrent reduces to a constant minimum which is
$e\Delta_{0}/\hbar$ for armchair edges and vanishingly small for
smooth edges. We further analyze the product $I_cR_N$ ($R_N$ being
the normal resistance of the SQPC) representing the strength of
the Josephson coupling, and the current-phase relations as a
function of $W/L$. While the product $I_cR_N$ has a constant value
of $2.08\Delta_{0}/e$ for a wide contact $W/L\gg1$ irrespective of
the edges type \cite{beenakker06-2}, in narrow junctions the
Josephson coupling value depends on the type of the edges. By
decreasing the width in smooth case, the product $I_cR_N$
decreases monotonically and reach a minimum $(\pi/2)\Delta_{0}/e$.
However, in armchair case it increases and tends to maximum of
$\pi\Delta_{0}/e$ for a narrow contact $W\lesssim L$. For graphene
SQPC the dependence of the current-phase relation on $W/L$ is also
different from an ordinary SQPC. For sufficiently narrow
nanoribbons, while the smooth-edge case have a sinusoidal
current-phase relation like the tunneling contacts, the
armchair-edge junction current-phase relation is
$I=I_{c}\sin(\phi/2)$ like an ordinary SQPC.
\par
Far from the Dirac point at high carrier concentrations, the
behavior of $I_c$ versus $W$ consists of a similar monotonic
variation and a series of peaks with distances inversely
proportional to the chemical potential $\mu$. We explain the
absence of the supercurrent quantization as a result of the
significant contribution of the evanescent modes in the
supercurrent which is a unique property of graphene monolayers.
Correspondingly the product $I_cR_N$ acquires an oscillatory
behavior as a function of $\mu W/\hbar \mathrm{v}_{F}$ with a
period $2\pi$. The oscillations for smooth and armchair edges have
a $\pi$ phase shift due to the difference in the zero-state
longitudinal momenta of two cases.
\par
For a junction with zigzag edges, because of the valley filtering
characteristic \cite{rycerz06} of the wave functions, situation is
drastically different from other two edges. We find that a zigzag
nanoribbon supports a half-integer quantization of the
supercurrent to $(n+1/2)4e\Delta_{0}/\hbar$ \cite{gz06} with a
current-phase relation identical to the ordinary SQPC.
\par
The paper is organized as follows. In next section we introduce
the model of the graphene SQPC and find the solution of the DBdG
equation. In Sec. 3 we describe the boundary conditions for
different edges and calculate the corresponding Josephson current.
Sec. 4 is devoted to analysis of the obtained results. Finally in
Sec. 5 our conclusion comes.
\begin{figure}
\centerline{\includegraphics [width=6cm]{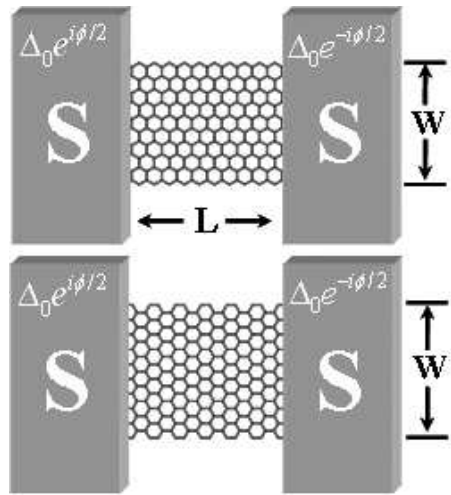}}
\caption{Schematic of the studied junction structures. Upper
(lower) panel shows a junction which has zigzag (armchair) edges.}
\label{fig:1} \label{gzfig1}
\end{figure}

\section{Model and basic equations}
\label{sec:2} The geometry of the studied Josephson junction is
shown schematically in figure \ref{gzfig1}. A ballistic graphene
strip (N) connects two wide superconducting regions (S) which are
produced by depositing of superconducting electrodes on top of the
graphene sheets. Recent experiments have verified the reality of
this model \cite{heersche,shailos}. We assume the interfaces
between the superconductors and the graphene strip are free of
defects and impurities. The superconducting parts are supposed to
be heavily doped such that the Fermi wavelength $\lambda_{F\,S}$
inside them is very smaller than the superconducting coherence
length $\xi$ and also the Fermi wavelength in the normal graphene
strip $\lambda_{F\,N}$. By the first condition mean field theory
of superconductivity will be justified and by second we can
neglect the spatial variation of $\Delta$ inside the
superconductors close to the normal-superconductor (NS)
interfaces. Thus the superconducting order parameter in the left
and right superconductors has the constant values
$\Delta=\Delta_{0}\exp(\pm i\phi/2)$, respectively, and vanishes
identically in N.
\par
The low-lying electronic states in graphene can be described by
the Dirac model as linear combinations of four zero-energy Bloch
functions with slow varying envelopes
$(\psi_{1},\psi_{2},\psi'_{1},\psi'_{2})$. Here $\psi_{1}$ and
$\psi_{2}$ are the wave function components on the two
inequivalent atomic sites in the honeycomb lattice usually
referred as pseudospin components. The primed and unprimed
functions denote the two inequivalent Dirac points $K$ and
$K^{\prime}$ in the band structure of graphene known as the two
valleys. The evolution of the envelope functions is governed by
the two dimensional Dirac Hamiltonian \cite{mele,ando,haldane}:
\begin{equation}
\hat{H}=-i\,\hbar \mathrm{v}_{F}(\tau_{0} \otimes
\sigma_{x}\,\partial_{x} +\tau_{z} \otimes
\mathbf{\sigma}_{y}\,\partial_{y}). \label{eq:1}
\end{equation}
The Pauli matrices $\sigma_{i}$ and $\tau_{i}$ act on the
sublattice and valley degrees of freedom, respectively (with
$\sigma_{0}$ and $\tau_{0}$ representing the $2 \times 2$ unit
matrices). The eigenfunctions of the Dirac hamiltonian are the
plane waves:
\begin{eqnarray}
\psi_{K}&=&e^{ikx+iqy}(1,e^{i\alpha},0,0),\nonumber\\
\psi_{K'}&=&e^{ikx+iqy}(0,0,1,e^{-i\alpha}), \label{eq:2}
\end{eqnarray}
in which the angle $\alpha=\arctan(q/k)$ indicates the propagation
direction.  By treating the pseudospin similar to the real spin,
one can easily see the pseudospin state $(1,e^{\pm i\alpha})$
aligns parallel with the propagation direction. This allows one to
introduce chirality, that is formally a projection of the
pseudospin on the direction of motion. The electron and hole have
positive and negative chirality, respectively \cite{katsnelson}.
In ballistic systems chirality of states does not change and this
conservation of chirality leads to important effects in scattering
phenomena, which Klein paradox is one of them
\cite{katsnelson,beenakker06-3}.

\par
The superconducting correlations between the electron-hole
excitations are described by the Dirac-Bogoliubov-de Gennes (DBdG)
equation for the electron and hole four-component wave functions
$u$ and $v$\cite{beenakker06-3}:
\begin{equation}
\left(\matrix{\hat{H}-\varepsilon-\mu & \hat{\Delta} \cr
\hat{\Delta}^{*} & \mu-\hat{\Theta} \hat{H}\hat{\Theta}
^{-1}-\varepsilon}\right)\left(\matrix{u\cr v}\right) = 0.
\label{eq:3}
\end{equation}
Here $\hat{\Delta}$ is the superconducting pair potential matrix
and $\hat{\Theta}$ is the time reversal operator. The excitation
energy $\varepsilon$ is measured relative to the Fermi energy
$\mu$. Each of the four blocks in Eq. \ref{eq:3} represents a $4
\times 4$ matrix, acting on two sublattices and two valleys
spaces. The matrix forms of the pair potential, and the time
reversal operators are:
\begin{equation}
\hat{\Delta}=\Delta\tau_{x}\otimes\sigma_{0}, \label{eq:4}
\end{equation}
\begin{equation}
\hat{\Theta}=\tau_{x}\otimes\sigma_{z}\cal{C}. \label{eq:5}
\end{equation}
in which $\cal{C}$ is the operator of complex conjugation.
\par
The singlet superconducting pairing always occurs between the
time-reversed states. So in the case of the Dirac fermions in
graphene, which the time-reversed states belong to the different
valleys, the superconducting pairing is between the particles with
different valleys as well as opposite spins and momenta. This
cause the time reversal symmetry to play a special role in the
Josephson effect compared to that in the phase coherent transport
in the normal state. Indeed the presence of a superconducting
electrode provides an intrinsic mechanism that couples phase
coherently the electronic states belonging to the opposite
valleys. So the dynamics of electrons is described by the full
two-valley Hamiltonian of graphene as indicated in the DBdG
equation.
\par
The phase difference $\phi$ between the order parameters drives a
supercurrent through the graphene strip which constitutes a weak
link between two superconductors. This Josephson supercurrent is
carried by the so called Andreev states which are formed in N
region due to the successive conversions of the electron-hole
excitations to each other (Andreev reflection) at the NS
boundaries. For a short junction of $L\ll\xi$ and at zero
temperature the Andreev bound (discrete) states with energies
$|\varepsilon|<\Delta_{0}$ have the main contribution to the
Josephson supercurrent \cite{beenakker91-1}. In this case we can
neglect the contribution from the continuous states above the
superconducting gap. We obtain the energies $\varepsilon(\phi)$ of
Andreev bound states by solving the DBdG equations with
appropriate boundary conditions which are described below. The
Josephson current is then obtained from the formula
\begin{equation}
I=\frac{2e}{\hbar}\frac{\delta F}{\delta \phi}, \label{eq:6}
\end{equation}
between the Josephson current and the derivative of the free
energy $F$ with respect to the phase difference.
\par
Inside N the solutions of the DBdG equation are independent
electron and hole-like wave functions which are classified by the
2-dimensional wave vector ${\mathbf{k}}\equiv(k,q)$ with the
energy-momentum relation $\varepsilon =\hbar
\mathrm{v}_{F}|\mathbf{k}|$. For a finite width $W$ the transverse
momentum $q$ is quantized by imposing the boundary conditions at
the edges. These transverse boundary conditions are different for
the different edge types which a graphene strip can have.
\par
The solutions inside the superconductors are rather mixed
electron-hole excitations, called Dirac-Bogoliubov quasiparticles.
The interfaces scatter the particles between the two neighboring
regions. However, the excitations inside N with energy inside the
gap cannot enter the superconductors and two kinds of processes,
Andreev and normal reflection can occur for them. In general to
match the solutions inside the different regions, one needs the
scattering matrices of the interfaces. But assuming the ideal NS
boundaries electron-hole scattering can be described by a
longitudinal boundary condition between the electron and hole wave
functions which reads \cite{beenakker06-2},
\begin{equation}
u=\tau_{x}\otimes e^{-i\varphi+i\beta\mathbf{n}\cdot\vec{\sigma}}\, v,\\
\,\,\,\,\beta=\arccos(\varepsilon/\Delta_{0})\\
\label{eq:7}
\end{equation}
where $\varphi$ is the phase of $\Delta$ in S and $\mathbf{n}$ is
the unit vector perpendicular to the interface pointing from N to
S.
\par
Assuming elastic processes, scattering at the interfaces does not
change the transverse momentum $q$ and there is not any scattering
between the different modes. So the Andreev states will be linear
combinations of the wave functions from the same mode $n$ with
certain transverse momentum $q_{n}$. For each type of the edges by
imposing the longitudinal boundary conditions at the interfaces to
a general linear combination of the electron and hole states of
mode $n$ we reach to an equation for the Andreev energies
$\varepsilon_n(\phi)$. Throughout the paper we will consider the
limit $\Delta_{0}\ll \mu$ where the retro-Andreev reflection
dominates \cite{beenakker06-3}. In this regime the longitudinal
momenta of the electron and hole $k_{n}(\pm \varepsilon)$ can be
approximated with $k_{n}(0)$ independent of the excitation energy
\cite{beenakker06-2}.

\section{Josephson current of the nanoribbons with different edges}
\label{sec:3}
\subsection{Smooth edges}
A graphene strip with smooth edges corresponds to an infinite mass
confinement which is described by the following boundary
conditions \cite{berry},
\begin{equation}
\psi_{1}(0)=-\psi_{2}(0), \,\, \psi_{1}(W)=\psi_{2}(W).
\label{eq:8}
\end{equation}
These conditions do not mix the two valleys and we can treat them
independently. By this conditions the transverse momentum is
obtained to have the quantized values
$q_{n}=(n+\frac{1}{2})\pi/W$. We note that this quantization
condition for the Dirac particles confined by an infinite mass
differs from the corresponding one for the ordinary normal
electrons by an offset of $1/2$. This originates from the Berry
phase $\phi_{B}=\pi$ in monolayer graphene which cause $\pi$ phase
shift in the boundary conditions as well \cite{zhang05,berry}.
Considering a linear combination of the plane wave solutions (Eq.
\ref{eq:2}) with opposite transverse momenta and applying the
boundary conditions we can find the modes inside the strip. The
electron modes for each valley are indicated by the wave functions
of the form:
\begin{equation}
u_{k}=e^{ikx}[\sin(qy-\alpha/2),\sin(qy+\alpha/2)]. \label{eq:9}
\end{equation}
The hole wave functions $v_{k}$ for each particular mode are the
time reversed of the corresponding electron mode. Andreev bound
states for each mode will be a linear combination of $u_{\pm k}$
and $v_{\pm k}$:
\begin{equation}
\Psi_{A}=\left(\matrix{u_{A}\cr
v_{A}}\right)=\left(\matrix{au_{k}+bu_{-k}\cr
a'v_{k}+b'v_{-k}}\right). \label{eq:10}
\end{equation}
To form an Andreev state this general form should satisfy the
longitudinal boundary conditions (Eq. \ref{eq:7}) at the NS
interfaces ($x=0, L$). These conditions result in four independent
homogeneous equations for four coefficients. By solving these
equations we obtain the quantized Andreev energy as
\begin{equation}
\varepsilon_{n}(\phi)=\Delta_{0}\sqrt{1-t_{n}\sin^{2}(\phi/2)},\\
\label{eq:11}
\end{equation}
with $t_{n}=[1+(q_{n}/k_{n})^{2}\sin^{2}(k_{n}L)]^{-1}$.
Comparison with Ref. \cite{beenakker06-1} shows that $t_{n}$ is
indeed the transmission probability for a ballistic smooth-edge
strip of graphene between two heavily doped electrodes in the
normal state. This type of relation between the Andreev energies
and normal-state transmission probabilities $t_{n}$ is a universal
form in the short junction limit ($\xi\gg L$)
\cite{beenakker91-1}. By means of this method we can also find the
transparency of each mode and then calculate the normal state
conductance ($G_{N}$) using the Landauer-Buttiker formula at zero
temperature:
\begin{equation}
G_{N}=R_{N}^{-1}=4 \frac{e^{2}}{h}\sum_{n}{t_{n}}.\\
\label{eq:12}
\end{equation}
The factor 4 at above relation accounts for the two-fold spin and
valley degeneracies. Now from the Andreev energies and also Eq.
\ref{eq:6} the Josephson current at zero temperature can be found.
\par
Let us concentrate on the transmission coefficients and their
dependence on the size and geometry of the junction. We note that
in general both types of oscillatory (real $k_{n}$) and evanescent
(imaginary $k_{n}$) modes participate in the transport.  The
number of propagating modes is limited by the width as compared to
the Fermi wavelength. Indeed only the modes with $n\leq N_{W}=\mu
W/\pi\hbar\mathrm{v}_{F}-1/2$ can propagate and all the upper
states are evanescent. Such a distinction between the modes exists
in an ordinary QPC but with an important difference. In the QPC
the transparencies depend on the longitudinal kinetic energy
$E^{\parallel}_{n}=\hbar^{2}k_{n}^{2}/2m$ as
$t_{n}=1/(1+e^{-E_{n}^{\parallel}/\zeta})$, in which $\zeta$ is a
function of the geometry and size of the channel
\cite{glazman,buttiker}. Because of the quadratic form of the
dispersion relation and high Fermi energies in comparison with
$\zeta$, we have $|E_{n}^{\parallel}/\zeta|\gg 1$ for all modes.
Consequently all the propagating modes are completely transmitted
and the evanescent modes have very small transparencies. But since
in graphene very small Fermi energies are accessible and the
energy depends linearly on the momentum the evanescent modes can
have a significant contribution in the transport.  In particular
near the Dirac point $\mu L/\hbar \mathrm{v}_{F}\ll 1$, the number
of propagating modes are very small and the evanescent modes have
the main contribution in the transport. This is the unique
property of the graphene Josephson junctions in which very small
Fermi energies are accessible. In this limit the transmission
coefficients through ballistic graphene are approximately,
\begin{equation}
t_{n}=\cosh^{-2}(q_{n}L), \label{13}
\end{equation}
which have the same form as transmission eigenvalues of the
diffusive transport through a disordered contact \cite{stone}.

\subsection{Armchair edges}
From Fig. \ref{gzfig1} (lower panel) one may see that at the
armchair edges the termination consists of a line of dimers.
Therefore the wave function amplitude vanishes on both sublattices
at the two edges. This leads to the boundary conditions,
\begin{equation}
\psi_{1}=\psi'_{1},\,\,\,\,\,\psi_{2}=\psi'_{2} \label{14}
\end{equation}
at the both edges. In contrast to the smooth edges, armchair edges
mix the two valleys and causes scattering of a wave propagating on
a certain valley to the opposite valley. The boundary conditions
leads to the quantized transverse momenta $q_{n}=n\pi/W$ similar
to the usual confinement of normal electrons. Due to the valley
changing property of these edges, the electron wave function of
each mode inside the nanoribbon is a mixture of the two valleys
with the form
\begin{equation}
u=e^{ikx}[e^{iqy}(1,e^{i\alpha}),
e^{-iqy}(1,e^{i\alpha})].\\
\label{eq:15}
\end{equation}
This wave function has the same chirality for both valleys
contributions, as a result of the symmetric boundary conditions.
Also one should note that in contrast to the smooth edges, the
wave functions $u(q)$ and $u(-q)$ represent distinct solutions in
which the components of the two valleys are interchanged. As a
result the lowest mode $n=0$ has not the two fold valley
degeneracy of the higher $n=1,2,...$ modes. This zero mode with
zero transverse momentum can exist inside the channel even if its
widths becomes smaller than the Fermi wavelength. Like the case of
the smooth edges the Andreev state wave function corresponding to
a given mode $n$ is the combination of $u_{\pm k}$ and $v_{\pm
k}$. But there is a difference with the smooth case. This time
$u_{A}$ and $v_{A}$ have four nonzero components. Applying the
longitudinal boundary conditions leads to 8 equations for the
coefficients $a$,$a'$,$b$,$b'$ in Eq. \ref{eq:10}. However because
of the symmetry between the contributions from the two valleys,
only 4 of them are independent. We find that the Andreev energies
have the same form as in the smooth strip (Eq. \ref{eq:11}). So we
would expect to reach similar results for $I_{c}$ and $G_{N}$ as
well.
\subsection{Zigzag edges}
We now consider the case of the zigzag edges. It is clear from
Fig. \ref{gzfig1} (upper panel) that atoms at zigzag edges belong
to different sublattices. The boundary conditions are
$\psi_{1}=\psi'_{1}=0$ at $y=0$ and $\psi_{2}=\psi'_{2}=0$ at
$y=W$ \cite{brey06}, which lead to the transcendental relation,
\begin{equation}
\sin(qW)/qW=\pm \hbar \mathrm{v}_{F} /(\mu+\varepsilon) W,
\label{eq:16}
\end{equation} for the transverse momentum $q$.
In contrast to the previous cases this relation couples $q$ to the
excitation energy $\varepsilon$. Of course in the regime of our
interest ($|\varepsilon|<\Delta_0\ll \mu$) we can neglect
$\varepsilon$ at Eq. \ref{eq:16}. This equation has a finite
number of solutions depending on the value of $\mu W/\hbar
\mathrm{v}_{F}$. Indeed all possible values of the transverse
momenta are smaller than $\mu/\hbar \mathrm{v}_{F}$ and thus the
longitudinal momenta always have real values. This means that all
the modes inside the nanoribbon with zigzag edges are propagating.
For $\mu W/\hbar \mathrm{v}_{F}<1$ there is an imaginary solution
$q$ labeling an evanescent mode in the $y$-direction.  In the
interval $1<\mu W/\hbar \mathrm{v}_{F}<3\pi/2$, the transcendental
relation has a single oscillatory solution. For larger $\mu
W/\hbar \mathrm{v}_{F}$ the number of the solutions increases by
two whenever the width is increased by a half of the Fermi
wavelength $\lambda_{F\,N}=h\mathrm{v}_{F}/\mu$. We classify the
wave functions to two groups according to $\pm$ sings in the
transcendental relation. To find the wave functions we consider
the general combination of the plane wave solutions:
\begin{equation}
\psi=A\psi^{q}_{K}+B\psi^{-q}_{K}+A'\psi^{q}_{K'}+B'\psi^{-q}_{K'}.\\
\label{eq:17}
\end{equation}
The boundary condition at $y=0$ leads to $A+B=A'+B'=0$. So the
wave function must be as below:
\begin{equation}
\psi=e^{ikx}\left(\matrix{A\sin(qy)\cr A\sin(qy+\alpha)\cr A'\sin(qy)\cr A'\sin(qy-\alpha)}\right).\\
\label{eq:18}
\end{equation}
Now to satisfy the second condition at $y=W$ there are two ways.
We can have either $\sin(qW+\alpha)=0$ and $A'=0$ or
$\sin(qW-\alpha)=0$ and $A=0$. We consider the solutions belonging
to the group with $\sin(qW)/qW=+\hbar \mathrm{v}_{F}/\mu W $. If
$k$ is positive the first way for satisfying the boundary
conditions will be possible and if $k$ is negative then the second
way is acceptable. So the electronic wave functions of the first
group ($+$ sign) have the forms,
\begin{eqnarray}
u^{+}=e^{ikx}[\sin(qy),\sin(qy+\alpha),0,0],\nonumber\\
u^{-}=e^{-ikx}[0,0,\sin(qy),\sin(qy-\alpha)], \label{eq:19}
\end{eqnarray}
which, respectively, describe a right-going wave on the valley $K$
and a left-going wave in the other valley $K^{\prime}$. It can be
easily seen that the wave functions of the second group ($-$ sign)
are obtained from the above functions by the replacement
$k\rightarrow-k$. We note that for each mode the zigzag nanoribbon
operates as a valley filter for the waves with mixed valley
components \cite{rycerz06}. Since the valley index and $q$ are
conserved upon a normal reflection, the filtering property
prevents the electrons from such scattering. Consequently the wave
function of each bound state in N will be a combination of $u^{+
(-)}$ and $v^{-(+)}$. From the longitudinal condition (Eq.
\ref{eq:7}) we obtain that the energy of the Andreev state does
not depend on the transverse mode $n$ and is given by,
\begin{equation}
\varepsilon=\Delta_{0}\cos(\phi/2),\label{eq:20}
\end{equation}
which is the same as the Andreev energies of an ordinary SQPC
\cite{beenakker91-1}.

\section{Results and discussion}
\label{sec:4}
\par
\begin{figure}
\centerline{\includegraphics [width=6cm]{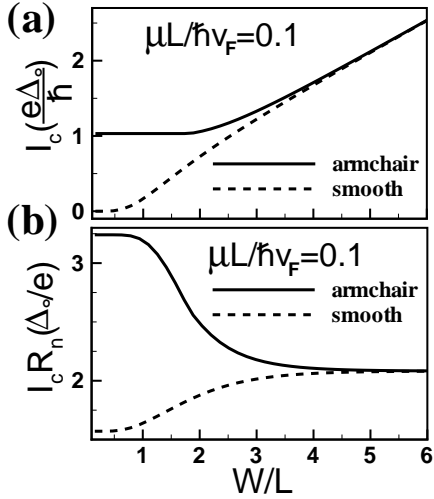}} \caption{The
critical current (a) and the product $I_{c}R_{N}$ (b) dependence
on the relative width of junction at a typical small value $\mu
L/\hbar \mathrm{v}_{F}=0.1$ for smooth and armchair strips.}
\label{gzfig2}
\end{figure}

Let us first consider the behavior of the Josephson current near
the Dirac point when $\mu L/\hbar \mathrm{v}_{F} \ll 1$. Fig.
\ref{gzfig2} shows the critical current and the product
$I_{c}R_{N}$ variations with $W/L$ for the smooth and armchair
edges at a typical small Fermi energy $\mu L/\hbar
\mathrm{v}_{F}=0.1$. We see $I_{c}$ has a linear dependence on
$W/L$ at large $W/L$ (\ref{gzfig2}a) which is the result of
diffusive-like transport in the ballistic graphene
\cite{beenakker06-2}. By decreasing $W/L$, the transmission
probabilities of the modes through N decrease and the critical
current shows a monotonic decrease without any quantization. For a
narrow strip $W\lesssim L$ the evanescent modes transparencies are
vanishingly small and $I_{c}$ reduces to a constant minimum value.
For the smooth edges there is no propagating mode when
$W<(\pi/2)\hbar \mathrm{v}_{F}/\mu$ and the minimum supercurrent
is vanishingly small. However in the case of the armchair edges a
lowest nondegenerate mode with $q=0$ always can propagate
irrespective of the width of the junction. This zero mode results
in a bound state carrying a nonzero residual supercurrent
$I_{c}=e\Delta_{0}/\hbar$ for $W\lesssim L$. (Fig. \ref{gzfig2}a)
\par
For a wide contact the normal state conductance $G_{N}$ has the
same linear dependence on $W/L$ as $I_{c}$ and thus the product
$I_{c}R_{N}$ reaches a constant value $2.08\Delta_{0}/e$ in the
limit $W\gg L$(see Fig. \ref{gzfig2}b) for both types of the
edges. Decreasing the width the difference between the Josephson
coupling constants of the junctions with two different edges
appears. For smooth (armchair) edges $I_{c}R_{N}$ decreases
(increases) monotonically  with lowering $W/L$ and reaches to a
minimum (maximum) of $(1/2)\pi\Delta_{0}/e$ ($\pi\Delta_{0}/e$)
for a narrow junction of $W\lesssim L$ (see Fig. \ref{gzfig2}b).

\par
\begin{figure}
\centerline{\includegraphics [width=6cm]{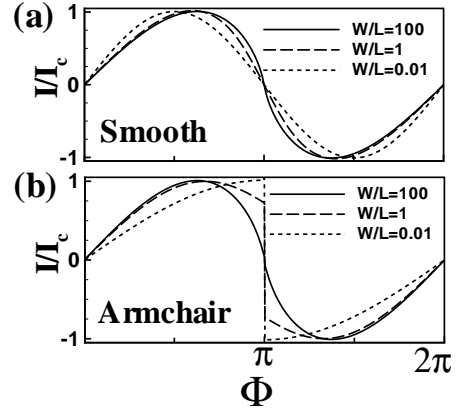}} \caption{The
current-phase relation for junctions with smooth and armchair
edges at different $W/L$ values for $\mu L/\hbar
\mathrm{v}_{F}=0.1$.} \label{gzfig3}
\end{figure}

As indicated in Fig. \ref{gzfig3} the current-phase relation of
finite-width junction differs significantly for the armchair and
smooth cases specially for very narrow strips. However at the
limit of very wide junctions $W\gg L$ for both types, the
current-phase relation is like a diffusive metal junction
\cite{beenakker91-2,kulik1}:
\begin{equation}
I(\phi)=\frac{e\Delta_{0}}{\hbar}\frac{2W}{\pi L}\cos(\phi/2)\mathrm{arctanh}(\sin(\phi/2)).\\
\label{eq:21}
\end{equation}
On the other hand a very narrow strip $W\ll L$ with smooth edges
has a sinusoidal current-phase relation $I(\phi)=I_{c}\sin(\phi)$
similar to a tunnel Josephson junction. This is expected as in
this limit the transmission probabilities are very small and the
graphene ribbon behaves as a tunnel junction. For a narrow
armchair ribbon the current-phase relation reads
\begin{equation}
I =
I_{c}\frac{\cos(\phi/2)}{|\cos(\phi/2)|}\sin(\phi/2)\\
\label{eq:22}
\end{equation}
which is the characteristic relation for the SQPC. This is also
understandable because even in a very narrow armchair strip there
is a propagating mode with complete transmission which makes it
similar to the SQPC. The current-phase relation of smooth and
armchair junctions are shown in Fig. \ref{gzfig3} for different
$W/L$.

\par
\begin{figure}
\centerline{\includegraphics [width=6cm]{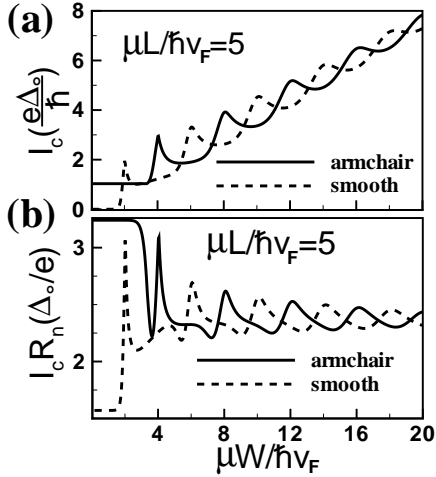}} \caption{The
critical current (a) and the product $I_{c}R_{N}$ (b) dependence
on the relative width of junction at a typical large value $\mu
L/\hbar \mathrm{v}_{F}=5$ for smooth and armchair strips.}
\label{gzfig4}
\end{figure}

Now let us consider the Josephson current at a higher Fermi energy
far from the Dirac point. Fig. \ref{gzfig4}a and \ref{gzfig4}b
show the dependence of $I_{c}$ and $I_{c}R_{N}$ on the junction
width, respectively, for $\mu L/\hbar \mathrm{v}_{F}=5$. In this
regime both the propagating and evanescent modes contribute to the
supercurrent. Roughly speaking the amplitude of the contribution
from the evanescent modes is proportional to $W/L$, however, the
propagating modes contribution is proportional to their number $
N_{W}\sim\mu W/\pi\hbar \mathrm{v}_{F}$. When $\mu L/\hbar
\mathrm{v}_{F}\ll 1$, the evanescent modes play the main role in
the transport and thus the transport is diffusive-like and both
the critical current and conductance are proportional inversely to
length of the junction $L$. On the other hand, when $\mu L/\hbar
\mathrm{v}_{F}\gg 1$ the propagating modes have the major
contribution. This cause that the envelope of the critical current
depends linearly on $\mu W/\hbar \mathrm{v}_{F}$. In addition to
this overall increase with $\mu W/\hbar \mathrm{v}_{F}$ the
critical current undergoes a series of peaks which become smoother
by increasing $\mu W/\hbar \mathrm{v}_{F}$. Each peak (jump)
signals addition of a new propagating mode in the transport when
the width increases by a Fermi wavelength. Both the critical
current and the product $I_{c}R_{N}$ have an oscillatory behavior
versus $\mu W/\hbar \mathrm{v}_{F}$ with a period of $2\pi$ (Fig.
\ref{gzfig4}). We should note that due to a nonzero transverse
momentum of the lowest mode ($n=0$) in the junction with smooth
edges (while the zero mode of the armchair-edge nanoribbon has
zero transverse momentum), the oscillations are shifted by half a
period with respect to those of the armchair edges case. Also
comparing the results of Fig. \ref{gzfig2} and \ref{gzfig4}, we
see that the limiting values of $I_{c}$ and ${I_{c}R_{N}}$ at
$W\lesssim L$ is always the same and does not depends on the value
of the chemical potential $\mu$. Therefore we find that for these
two types of the edges there is no sharp quantization and stepwise
variation with width for $I_{c}$ and $G_N$. This is again a unique
property of graphene with smooth and armchair edges in which the
evanescent modes can contribute significantly in the transport.

\par
\begin{figure}
\centerline{\includegraphics [width=6cm]{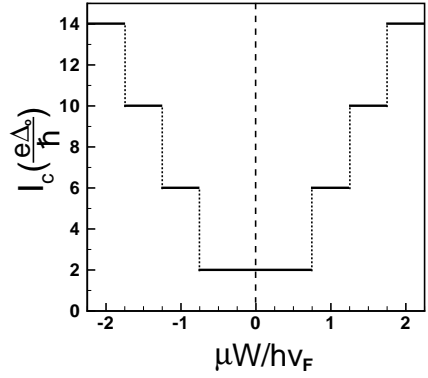}} \caption{The
critical current \textit {vs.} relative width of junction which
has zigzag edges.} \label{gzfig5}
\end{figure}

Finally let us analyze the Josephson supercurrent for the zigzag
ribbon.  As we argued in section \ref{sec:3} in this case the
valley filtering nature of the wave functions in N prevents the
normal reflections at NS interfaces. In addition, because of the
transcendental relation of the transverse momenta which confines
their values by the Fermi energy $\mu$, there is no evanescent
mode with imaginary $k$ and the situation is similar to an
ordinary SQPC. The critical supercurrent shows a step-wise
variation with $\mu W/h{\rm v}_{F}$ but with the following
important anomalies (see Fig. \ref{gzfig5}). In contrast to an
ordinary SQPC \cite{beenakker91-1,furusaki91}, the width of the
first step ($3/4$) is bigger than that of the higher steps
($1/2$). The extra width is the contribution of the single
$y$-directional evanescent mode (imaginary $q$) for $W<\hbar {\rm
v}_{F}/\mu $. Also the height of the first step is $1/2$ of the
height of the higher steps which itself is four times bigger than
$e\Delta_{0}/\hbar$ the supercurrent quantum in an ordinary SQPC.
Therefore the supercurrent through a zigzag graphene nanoribbon is
half-integer quantized to $(n+1/2)4e\Delta_{0}/\hbar$. The effect
resembles the conductance quantization in graphene strips with
zigzag edges \cite{peres06} and also the half-integer Hall effect
\cite{novoselov05,zhang05,gusynin05} in monolayer structures. From
expression of the Andreev energies for the zigzag junction (Eq.
\ref{eq:20}) we find that the current-phase relation is given by
Eq. \ref{eq:22} similar to an ordinary SQPC with length  smaller
than $\xi$.

\section{Conclusion }\label{sec:5}
In conclusion we have investigated the Josephson effect in a short
graphene nanoribbons of width $W$ and length $L\ll \xi_0$
connecting two heavily doped superconducting electrodes. Within a
Dirac-Bogoliubov-de Gennes formalism, we have found that the
variation of the Josephson supercurrent versus the width $W$ is
drastically different for different types of edges. In the smooth
and armchair nanoribbons with low concentration of the carriers
the critical supercurrent $I_c$ decreases monotonically by
decreasing $W/L$. For a narrow strip $W\lesssim L$ with armchair
and smooth edges, $I_c$ takes constant minimums
$e\Delta_{0}/\hbar$ and 0, respectively. The Josephson coupling
strength given by the product $I_cR_N$ has been found to have a
minimum $(1/2)\pi\Delta_0/e$ (maximum $ \pi\Delta_0/e$) for a
narrow junction $W\lesssim L$ with smooth (armchair) edges and
increases (decreases) monotonically with $W/L$ to reach the wide
junction value $2.08\Delta_{0}/e$. We have also seen the different
dependence of the current-phase relation on $W/L$ for junctions
with the two different edge types.
\par
For the higher concentration of the carriers, the overall
monotonic dependence of $I_c$ acquires a series of peaks with
distances inversely proportional to the chemical potential $\mu$.
Correspondingly the product $I_{c}R_{N}$ shows an oscillatory
behavior with $\mu W/\hbar \mathrm{v}_{F}$ with a period of
$2\pi$. The oscillations for smooth and armchair edges are phase
shifted by $\pi$.
\par
For the zigzag edges the results are quite different due to the
special quantization relation of transverse momenta and the valley
filtering property of the electron wave functions. We have found a
step-wise variation of $I_{c}$ versus $\mu W/h\mathrm{v}_{F}$,
implying a half-integer quantization of the supercurrent to
$(n+1/2)4e\Delta_{0}/\hbar$. The current-phase relation is found
to be similar to an ordinary superconducting quantum point
contact.

\section*{ Acknowledgments}
One of the authors (M. Z.) thanks A. Brataas for his hospitality
and support at the Center for Advanced Study, Oslo where parts of
this work were done.

\end{document}